\documentclass[manuscript]{aastex62}

\usepackage{lineno}
\usepackage{comment}
\usepackage{amsmath}
\usepackage{amssymb}
\usepackage{booktabs}
\received{xx}
\revised{xx}
\accepted{xx}
\begin{document}
\shortauthors{Hou et al.}


\title{Contribution of Magnetic Reconnection Events to Energy Dissipation in Magnetosheath Turbulence}

\correspondingauthor{Jiansen He}
\email{jshept@pku.edu.cn}

\author{Chuanpeng Hou}
\affiliation{School of Earth and Space Sciences, Peking University \\
Beijing, 100871, China; E-mail: jshept@pku.edu.cn}

\author{Jiansen He}
\affiliation{School of Earth and Space Sciences, Peking University \\
Beijing, 100871, China; E-mail: jshept@pku.edu.cn}

\author{Xingyu ZHU}
\affiliation{School of Earth and Space Sciences, Peking University \\
Beijing, 100871, China; E-mail: jshept@pku.edu.cn}

\author{Ying Wang}
\affiliation{School of Earth and Space Sciences, Peking University \\
Beijing, 100871, China; E-mail: jshept@pku.edu.cn}

\begin{abstract}

By analyzing the magnetosheath measurements from MMS, we obtain the statistical results for the contribution of magnetic reconnection (MR) events at electron scales to the energy dissipation of coherent structures. The Partial Variance of Increments (PVI) method is employed to find coherent structures in the magnetic field data. The current sheet structures with reversal of magnetic field components are further selected. We consider the following criteria to identify the MR events, such as current sheet with magnetic field reversal, significant energy dissipation, and evident electron outflow velocity. Statistically, for most MR events, their PVI values are larger than that of other types of coherent structures, and their energy dissipations are also stronger than that of others. However, due to the relatively small proportion of MR events, their contribution to coherent structures' energy dissipation is relatively trivial. If taken into account the dissipation of non-coherent structures, the MR's contribution to energy dissipation would be less. Hence, we suggest that MR events, though have strong dissipation locally, are not the major contributor to the energy dissipation in the magnetosheath. After analyzing the features of non-MR current sheets, we propose that non-MR current sheets are mainly coherent structures inherent to kinetic Alfv\'en fluctuations.
\end{abstract}


\section{Introduction}
	\label{S-Introduction}

Compared with the upstream solar wind, plasmas in the magnetosheath have higher temperature, higher density magnetic magnitude, and stronger turbulence. The strong magnetosheath turbulence, which is resultant from the amplification of solar wind turbulence \citep{schwartz1996low} and excitation of sheath plasma instability \citep{gary1993ion}, is an ideal environment to observe various types of energy dissipation phenomena and understand the comprehensive framework of turbulence dissipation mechanisms. Coherent structures usually appear intermittently in turbulence. Turbulence intermittency have been widely studied in theory \citep{burlaga1991intermittent, marsch1997intermittency}), simulations \citep{greco2008intermittent, wan2016intermittency} and observations \citep{marsch1997intermittency, osman2014magnetic, chasapis2017electron}. For example, tangential-discontinuity associated current sheets, which is a kind of coherent structures, can be strongly dissipated and hence heat local plasmas prominently \citep{wang2013intermittent}. By contrast, the rotational-discontinuity-associated current sheet does not contribute significantly to local plasma heating since it propagates through plasmas instead of co-travels with the same patch of plasma over time \citep{wang2013intermittent,zhang2015occurrence}. How are the intermittent coherent structures of various kinds related to turbulence dissipation and plasma heating? It is desirable to understand various specific processes in the comprehensive scenario of turbulence energy dissipation by quantifying and comparing contributions from different turbulence dissipation mechanisms.
	

Specifically, the turbulence dissipation mechanism can be classified into the dissipation of coherent structures and dissipation of waves. One class of coherent structure dissipation refer to magnetic reconnection (MR) \citep{drake2003formation,retino2007situ,shay2014electron,Fu2016,wang2016coalescence,ergun2020observations} and non-MR coherent structures \citep{alexandrova2008alfven, wang2013intermittent, wan2016intermittency,chasapis2018situ,huang2018observations, wang2019vortex}. Dissipation associated with waves includes resonance interactions between waves and particles, such as landau resonance \citep{chen2019evidence} and cyclotron resonance \citep{hollweg2002generation,isenberg2019perpendicular}, and non-resonant random interactions between waves and particles \citep{chandran2010alfven,chandran2013stochastic}. Thanks to the MMS measurements of current density and electric field with high cadence and high quality, the dissipation rate on scales (i.e., dissipation rate spectrum) can be derived. The dissipation rate spectra can measure the strength of wave-particle interaction, the partition of dissipated energy among different species, and along with different directions \citep{he2019direct,he2020spectra}.
	
	
MR is a vital energy conversion process in the space environment \citep{priest1986magnetic}. In the observation data, physical quantities characteristic can be used to identify MRs \citep{gosling2007observations,phan2007evidence,zhou2017coalescence}, such as the magnetic field direction reverses, particles velocity increases, particles number density increases. The Magnetospheric Multiscale spacecraft (MMS), as characterized by its state-of-art high cadence of field and particle measurements, provides the possibility of studying electronic scale coherent structures \citep{burch2016magnetospheric}. The positive peak of $\mathbf{j}\cdot\mathbf{E'}$, which represents energy dissipation intensity \citep{sundkvist2007dissipation,he2019direct}, and the peak of the electric field in the direction parallel to the magnetic field are also features of MRs \citep{wilder2018role}. The velocity distribution of electrons within and near the MR diffusion region is usually non-gyrotropic. \citet{burch2016electron} reported an electron scale MR, whose electron velocity distribution was presented as a "crescent". \citet{wilder2018role} statistically analyzed MR in the magnetosheath and found that with the increase of the guiding field, the non-gyrotropy of electron velocity distribution decreases. MMS observations have also shown that there are electron scale MR events without an ion diffusion region in turbulent magnetosheath \citep{phan2018electron}. 
	 
 
In this work, we aim to use MMS data to identify the electron scale MR events, and quantify the contribution of MR to the dissipation of coherent structures as well as to the total dissipation of turbulence in the magnetosheath. In section \ref{sec2}, we introduce and employ the method of quantifying magnetic disturbances to identify intermittent coherent structures in turbulence. In section \ref{sec3}, we present the MMS data that we used. In section \ref{sec4}, we pick out MR events from the identified coherent structures according to a set of appropriate procedures. Quantitative statistical results of contributions from MR, non-MR coherent structures, and non-structure waves to the total turbulence dissipation are analyzed in section \ref{sec5}. The related conclusion is drawn in sections \ref{sec6}. 

\section{quantitative representation of magnetic disturbance } \label{sec2}
	
We apply the Partial Variance of Increments (PVI) method to quantify magnetic disturbance with the PVI index, and search for coherent structures with high dissipation. PVI method has been widely used to detect coherent structures \citep{greco2009statistical,chasapis2018situ}, and study turbulence intermittency \citep{greco2008intermittent}. In the solar wind, the PVI index is strongly correlated with temperature anisotropy and plasma heating \citep{osman2012kinetic}. For the magnetosheath turbulence at kinetic scales, the PVI index seems to show a positive correlation with the dissipation intensity \citep{chasapis2018situ}.

The formula of calculating PVI index reads as
\begin{eqnarray}
	\mathbf{PVI}(t_i,\tau) = \frac{|\mathbf{B}(t_i+\tau) - \mathbf{B}(t_i)|}{\sqrt{ <|\mathbf{B}(t_i+\tau) - \mathbf{B} (t_i)|^2>} }, \label{eq1}
\end{eqnarray} 
where $t_i$ represents the time, $ \tau $ represents the time lag between two time moments, and $<...> $ denotes an ensemble average.
In this paper, we focus on the electron-scale MR, so the time lag is selected to be the same as the electron inertial scale.

The result of applying the PVI method to magnetic data (Fig. \ref{fig1}a) is showed in Fig. \ref{fig1}b. By comparing the time series of magnetic field and PVI index, we can see that the stronger the magnetic field changes, the larger the PVI index appears. In Fig. \ref{fig1}f, we show the time series of $\mathbf{j}\cdot\mathbf{E'}$, which can be regarded as the dissipation rate of turbulent fields when $\mathbf{j}\cdot\mathbf{E'}>0$ at the kinetic scales. It can be seen that the peaks of $\mathbf{j}\cdot\mathbf{E'}$ do not always correspond to the peaks of the PVI index. The concurrence between the peaks of $\mathbf{j}\cdot\mathbf{E'}$ and PVI represents the dissipation caused by coherent structures. At other places where evident signals of $\mathbf{j}\cdot\mathbf{E'}$ are not associated with strong PVI, the dissipation is contributed by non-structure fluctuations, such as waves. Magnetic reconnection is one type of magnetic energy dissipation related to coherent structures. To statistically survey MR events, it would be convenient to first find out the places with the concurrence of PVI and $\mathbf{j}\cdot\mathbf{E'}$ peaks, and then diagnose the variable profiles carefully to identify the MR events finally.

\section{Overview of measurements by MMS}
 \label{sec3}

To determine MR's contribution to energy dissipation in the space plasma turbulence, we analyze the MMS measurements in the magnetosheath. We use data of MMS1 in the time interval from 06:33:44 to 11:08:40 UT on 2015-10-25. As an example, the time series measured by fast plasma instrument \citep{pollock2016fast}, fluxgate magnetometer \citep{russell2016magnetospheric} and electric field double probe instrument \citep{torbert2016fields} of MMS are shown in Fig. \ref{fig1}. The magnetic field, velocity, and electric field are full of strong fluctuations and sudden enhancements in intensity. The PVI index represents the strength of magnetic disturbances, and has been introduced in details in section \ref{sec2}. We use $\mathbf{j}=e(n_e\mathbf{v_e}-n_i\mathbf{v_i})$ to calculate electric current density. The work done by the electric field on plasma particle species ($\mathbf{j}\cdot \mathbf{E'}$) is calculated in the plasma rest reference frame ($\mathbf{E'}=\mathbf{E_{GSE}}+\mathbf{v\times B}$) to represent the energy conversion between fields and particles. The turbulence dissipation intensity can have sharp peaks sporadically distributed in the time and space domain. Thanks to the high temporal resolution of measurements from MMS satellites, we have the opportunity to study these disturbances and sharp peaks at electron scales. The contribution of different mechanisms to the total dissipation can be studied by classifying $\mathbf{j}\cdot\mathbf{E'}$ into different types according to the disturbances of the magnetic field and species' bulk velocity. 

\section{Method of Identifying Magnetic Reconnection Events} \label{sec4}

After setting a certain PVI threshold level, which is adjustable, we can count the number of coherent structures with the PVI index greater than the threshold level and the dissipation rate being significant. The coherent structures can be regarded as candidates for further identification of MR events. We perform a coordinate transformation to the magnetic field vectors and bulk velocity vectors of coherent structures based on the Minimum Variable Analysis (MVA) of the magnetic field vectors. Magnetic field vectors in the new LMN-coordinates have maximum, intermediate, and minimum variance in the L, M, and N directions. An encounter of a typical MR event is usually characterized with a depression of magnetic field strength ($\delta|{\mathbf{B}}|$), a reversal in magnetic field component ($B_L$), a peak of particle species' bulk velocity component (${v_L}$) and an evident dissipation intensity ($\mathbf{j}\cdot\mathbf{E'}$). In the following, we will elaborate on how to identify MR events as encountered by spacecraft statistically.

After obtaining the time series of PVI index based on MMS magnetic field data, we employ three analysis steps to each PVI index peak. First, we investigate every PVI peak to judge whether it corresponds to a $\mathbf{j}\cdot\mathbf{E'}$ peak. For the events with concurrence of PVI and $\mathbf{j}\cdot\mathbf{E'}$ peaks, we implement the MVA-based coordinate transformation to the data containing the peak of PVI index to obtain the magnetic field and velocity in the LMN coordinate system. Second, we calculate the relative difference of magnetic field strength between the PVI peak at the center and the ambient two sides, ($|\mathbf{B}|_{ambient}-|\mathbf{B}|_{PVI\_peak})/|\mathbf{B}|_{ambient}$. The ambient $|\mathbf{B}|$ is assigned with the value of larger $|\mathbf{B}|$ of the two sides. If $(|\mathbf{B}|_{ambient}-|\mathbf{B}|_{PVI\_peak})/|\mathbf{B}|_{ambient}>0.1$, we then mark it as the candidate of MR event. Thirdly, we check whether a reversal of $B_L$ happens across the PVI index peak and whether there is a peak in the electron bulk velocity component $v_L$. If the observational data containing the PVI index peak time satisfies the preceding three steps, then an MR event can be determined. Since MVA-based transformation depends on the time range, multiple time ranges are tested for MVA transformation to improve the MR events' recognition rate. Considering that the times of PVI index peak, $\mathbf{j}\cdot\mathbf{E'}$ peaks, zero of $ {B_L}$, and outflow velocity peak of an MR event do not usually match accurately to each other, so we allow them to have a small time deviation. In order to verify the feasibility of the identification method, we tested it with the catalog of MR events provided in \citet{wilder2018role}, and concluded that the method program successfully and automatically identified the MR events listed in the catalog. Therefore, the program of MR automatic identification method as introduced in this work is feasible, and the identified MR events as the output results can be credible.
 
According to the preceding processes, we identify an MR event from MMS1 data (Fig. \ref{fig1}). The time of MR occurs at about 11:07:46.60 on 2015-10-25, as shown in Fig. \ref{fig1}h. This MR has been reported by \citet{wilder2018role}. Our program can also identify other MR events that have not been reported before. Fig. \ref{fig2} shows an example of a newly identified MR event. By invoking the program, we survey the coherent structures and MR events based on MMS measurements from 06:33:44 to 11:08:40 on 2015-10-25. The result will be discussed in section \ref{sec5}.
  
 \section{Analysis of Energy Dissipation Contributions} \label{sec5}
 
By integrating the dissipation intensity, $\mathbf{j}\cdot \mathbf{E'}$, related to MR events and comparing it with the dissipation intensity of coherent structures, the contribution of the reconnection events to the dissipation of coherent structures can be determined. Table \ref{table_1} lists the number of coherent structures with strong dissipation, the number of MR events, and the corresponding dissipation intensities under the condition of different PVI index thresholds. As can be seen from Table \ref{table_1}, with the increase of PVI index thresholds, the number of coherent structures, the total dissipation intensity, the number of reconnection events, and the dissipation due to reconnection all decrease. According to the data in Table \ref{table_1}, we illustrate the histograms of proportions in terms of number counts and dissipation intensity in Fig. \ref{fig3}a.

With the increase of the PVI index threshold, the number of coherent structures and MRs gradually decreases. While the proportions of the number count and dissipation intensity of MR events to their counterparts of coherent structures increases. This increasing trend indicates that coherent structures with a larger PVI index threshold are more likely to be associated with magnetic reconnection. When the PVI index threshold equals to 3.5, the proportion of MR dissipation suddenly decreases, which may be due to the low number of MR events or due to the appearance of other strong coherent structures.

From Fig.~\ref{fig3}a, we know that the proportions of dissipation intensity contributed from the MR events to the dissipation amount of coherent structures did not exceed 15$\%$. This suggests that magnetic reconnection events contribute a small proportion to the total energy dissipation of coherent structures rather than acting as the primary contributor. If taking the dissipation of non-structures into account, the contribution of MR events would become even smaller, saying about 1$\%$. Under the same PVI index threshold, the proportion of MR dissipation (blue bar) is higher than that of the number of MR (black bar), indicating that the dissipation intensity per MR event is slightly higher than that per coherent structure. This suggests that reconnection is locally easier to achieve the dissipation of field energy than other coherent structures. Despite this, due to the small proportion of MR number count in the magnetosheath, the contribution to the total field-energy dissipation from MR events is small. 

In Fig.~\ref{fig3}b, we plot a tree diagram to illustrate intuitively the contribution ratios of different dissipation mechanisms to the total field energy dissipation (PVI index threshold is set to 1 for this tree diagram plot). As shown in Fig. \ref{fig3}b, the MR contribution is about 5$\%$ in the dissipation of coherent structures. Contribution from MRs will be lower if we consider the dissipation of non-coherent structures, e.g., wavelike fluctuations. So magnetic reconnection may not be the dominant mechanism of field-energy dissipation in the magnetosheath turbulence, which may be dissipated through various co-existing mechanisms.

Besides the MR-related current sheets, we also find many non-MR current sheets in the earth's magnetosheath. The reversal of the magnetic field also characterizes these current sheets, but their difference from the MR current sheets is the lack of peak velocity in the non-MR current sheets. To investigate the nature of these current sheets, we calculate the correlation between electron bulk velocity and magnetic field in the LMN coordinate system (see Fig. \ref{fig4}). In the L-direction, the peak of probability distribution of the correlation coefficient between $v_{eL}$ and $B_{L}$ (CC($v_{eL}, B_L)$) is close to 1. This indicates that there is a velocity shear in the L-direction across the non-MR current sheets. The correlated velocity and magnetic shears make the non-MR current sheet propagate through the plasmas like waves rather than stay static relative to plasmas. And there exist no plasma inflows to drive magnetic fluxes converging towards the current sheets. Based on the correlation between velocity and magnetic field as well as the width scale of the current sheets, we proposed that the non-MR current sheets may be caused by nonlinear steepening of kinetic Alfvén waves.

\section{Summary} \label{sec6}

\begin{figure}[htb!]
	\centerline{\includegraphics[width=13cm,clip=]{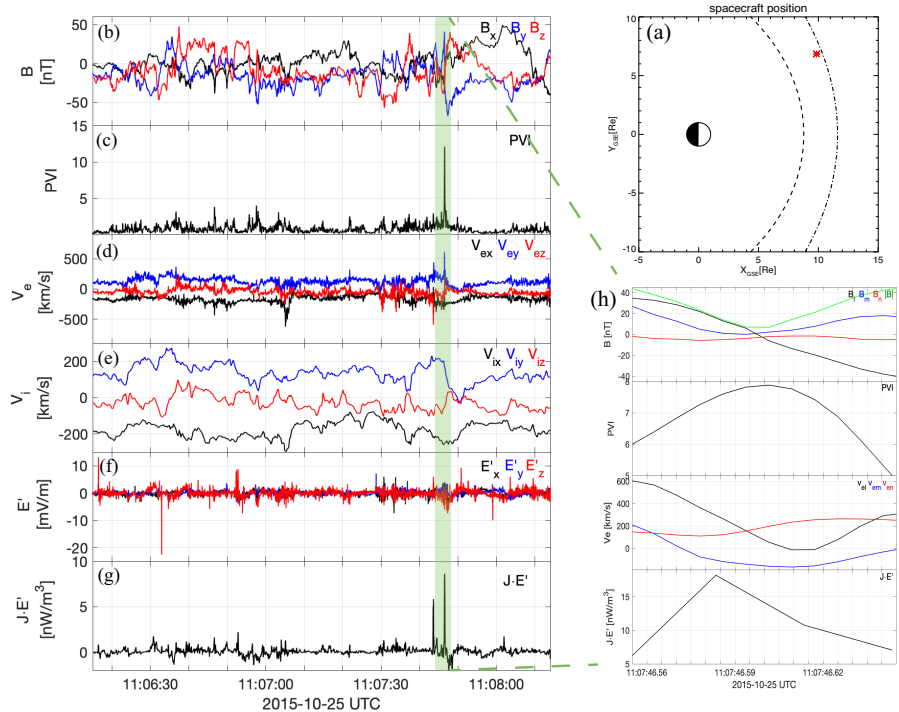}}
	\caption{(a) Position of the MMS1 spacecraft located in the magnetosheath. The bow shock (black dotted–dashed line) and magnetopause (black dashed line) are calculated by a 3D empirical model \citep{chao2002models}. (b) Magnetic field vectors in GSE coordinates. (c) PVI index series of the magnetic field. (d) Electron bulk velocity vectors in GSE coordinates. (e) Ion bulk velocity vectors in GSE coordinates. (f) Electric field vectors in the plasma reference reframe ($\mathbf{E'}= \mathbf{E_{GSE}}+\mathbf{v\times B}$). (g) Energy conversion rate (can be referred to as turbulence dissipation strength for turbulence study). (h) An example of MR picked out by the automatic identification program developed in this work.}
	\label{fig1}
\end{figure}

\begin{figure}[htb!]
	\centerline{\includegraphics[width=10cm,clip=]{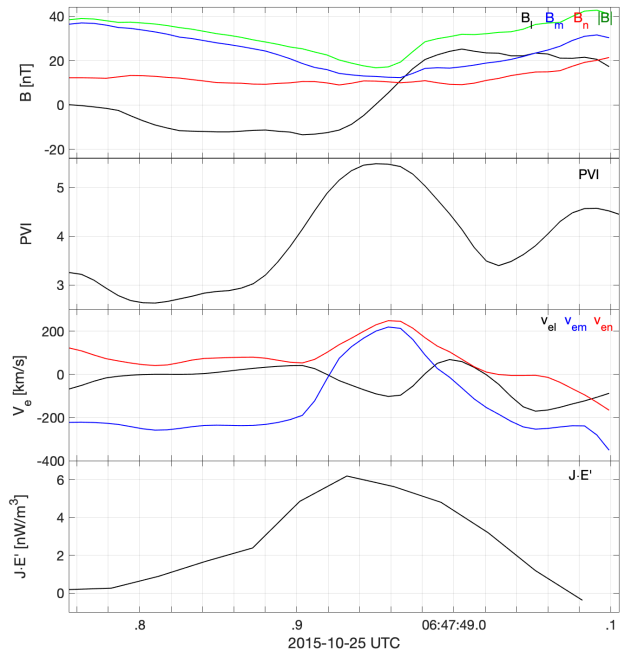}}
	\caption{ An example of newly identified MR event as picked out by our automatic identification program.}
	\label{fig2}
\end{figure}

\begin{figure}[htb!]
	\centerline{\includegraphics[width=15cm,clip=]{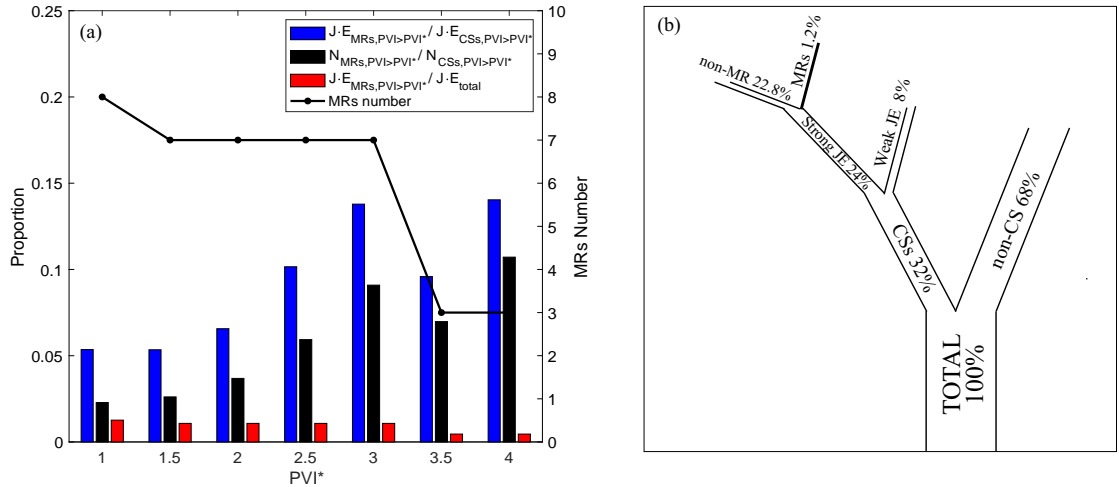}}
	\caption{(a) Statistical results about MRs and coherent structures in terms of number ratio and dissipation contribution under different PVI thresholds. The left axis represents three types of ratios: the percentage of field-energy dissipations between MRs and coherent structures (blue bars), the ratio of numbers between reconnection events and coherent structures (black bars), and the ratio of dissipations between reconnection events and total fluctuations (red bars). The right axis is used to mark the number of MR events (solid black line). (b) Tree diagram representing the proportion of field-energy dissipation contributed from different mechanisms: dissipation of non-coherent structures (e.g., kinetic waves) (68$\%$), dissipation of non-MR coherent structures (31$\%$), dissipation of MR events (1$\%$).}
	\label{fig3}
\end{figure}

\begin{figure}[htb!]
	\centerline{\includegraphics[width=10cm,clip=]{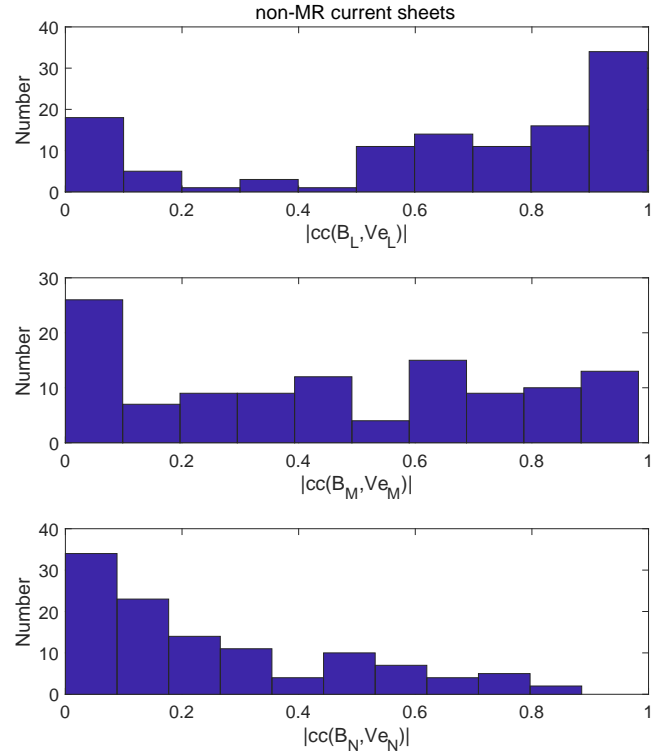}}
	\caption{Distribution of correlation coefficient (CC) between electron bulk velocity and magnetic field in LMN coordinate system. (a) Distribution of $|$CC$(B_{L}, v_{eL})|$ with the maximum number located in [0.9, 1.0]; (b) Distribution of $|$CC$(B_M, v_{eM})|$ with a rough even distirbution between [0,1]; (c) Distribution of $|$CC$(B_N, v_{eN})|$ with the maximum number located in [0.0, 0.1].}
	\label{fig4}
\end{figure}

We carry out a statistical analysis of the contribution of MR to the turbulence dissipation in the magnetosheath. We use the PVI method to identify the coherent structures with strong magnetic disturbance, and use the MVA-based coordinate transformation to analyze the sub-time-interval of vector sequences surrounding each PVI index peaks separately. We develop a program to identify MR automatically, which is proven to be feasible (see Fig. \ref{fig1}h and Fig. \ref{fig2}). To determine the reconnection events, we take into account the crucial features such as the reversal of magnetic field, decrease of magnetic magnitude, increase of electron velocity, and increase of dissipation intensity. Based on the thorough analysis of the identified MR events, coherent structures, as well as wavelike fluctuations, we draw the following conclusions.
 
 1. With the increase of the PVI index threshold, the contribution from MR to coherent structures, in terms of number proportion and dissipation proportion, generally increases. A larger PVI index is more likely to be associated with an MR event, and this association is consistent with the simulation work of \citet{servidio2011statistical}. 
 
 2. At different PVI index thresholds, the dissipation intensity of a single MR is slightly stronger. But due to the small proportion of MR, their contribution to the energy dissipation of coherent structures with strong dissipation is relatively trivial, with the proportion being less than 15$\%$. When the dissipation of non-coherent-structure fluctuations is taken into account, the contribution of MR will be even smaller (1$\%$). Therefore, we conclude that MRs may not be the dominant mechanism of turbulence dissipation in the magnetosheath.
 
 3. Besides the MRs, there are many non-MR current sheets in the magnetosheath. Electron bulk velocity and magnetic field of the non-MR current sheets in the L direction of LMN coordinate system show a good correlation, with the correlation coefficient being close to 1. Such $v-B$ correlation indicates that the non-MR current sheets may be related to the steepening of kinetic Alfvén waves.

\begin{table}[!t]  \caption{ Statistical Results of coherent structures with strong dissipation and MRs}

 \label{table_1}

 \begin{tabular}{lllll}  

\toprule   

  PVI* & Total Dissipation [$\frac{nW\cdot s}{m^3}$] & MR Dissipation  [$\frac{nW\cdot s}{m^3}$] (Proportion) & CSs Number & MR Number (Proportion)\\  

\midrule   

  1 & 202 & 10.8 (0.05)  & 350 & 8 (0.02)  \\  

  1.5 &  172 & 9.2(0.05) & 268 & 7 (0.03)  \\    

 2 & 140	 & 9.2 (0.07) & 190 & 7 (0.04)  \\

 2.5 & 90.8 & 9.2 (0.10) & 118 & 7 (0.06)  \\
 
 3 & 66.9	 & 9.2 (0.14) & 77 & 7 (0.09)  \\
 
 3.5 & 40.8	 & 3.9 (0.10) & 43 & 3  (0.07)   \\
 
 4 & 27.9 & 3.9 (0.14) & 28 & 3   (0.11)  \\

  \bottomrule  

\end{tabular}
$\mathbf{Note}$: PVI* represents the PVI index threshold. CSs stands for coherent structures. If considering the non-structures dissipation, the total energy dissipation is 856 $\rm{nW\cdot s/m^3}$.
\end{table}


\bigbreak

\noindent Acknowledgements:
This work is supported by NSFC under contracts 41874200, 41674171, and 41421003. The team is also supported by CNSA under contract Nos. D020301 and D020302. The authors are grateful to the teams of the MMS spacecraft for providing the data.


\begin{thebibliography}{}
\expandafter\ifx\csname natexlab\endcsname\relax\def\natexlab#1{#1}\fi
\providecommand{\url}[1]{\href{#1}{#1}}
\providecommand{\dodoi}[1]{doi:~\href{http://doi.org/#1}{\nolinkurl{#1}}}
\providecommand{\doeprint}[1]{\href{http://ascl.net/#1}{\nolinkurl{http://ascl.net/#1}}}
\providecommand{\doarXiv}[1]{\href{https://arxiv.org/abs/#1}{\nolinkurl{https://arxiv.org/abs/#1}}}

\bibitem[{Alexandrova \& Saur(2008)}]{alexandrova2008alfven}
Alexandrova, O., \& Saur, J. 2008, Geophysical research letters, 35

\bibitem[{Burch {et~al.}(2016{\natexlab{a}})Burch, Moore, Torbert, \&
  Giles}]{burch2016magnetospheric}
Burch, J., Moore, T., Torbert, R., \& Giles, B. 2016{\natexlab{a}}, Space
  Science Reviews, 199, 5

\bibitem[{Burch {et~al.}(2016{\natexlab{b}})Burch, Torbert, Phan, Chen, Moore,
  Ergun, Eastwood, Gershman, Cassak, Argall, {et~al.}}]{burch2016electron}
Burch, J., Torbert, R., Phan, T., {et~al.} 2016{\natexlab{b}}, Science, 352

\bibitem[{Burlaga(1991)}]{burlaga1991intermittent}
Burlaga, L. 1991, Journal of Geophysical Research: Space Physics, 96, 5847

\bibitem[{Chandran {et~al.}(2013)Chandran, Verscharen, Quataert, Kasper,
  Isenberg, \& Bourouaine}]{chandran2013stochastic}
Chandran, B., Verscharen, D., Quataert, E., {et~al.} 2013, The Astrophysical
  Journal, 776, 45

\bibitem[{Chandran(2010)}]{chandran2010alfven}
Chandran, B.~D. 2010, The Astrophysical Journal, 720, 548

\bibitem[{Chao {et~al.}(2002)Chao, Wu, Lin, Yang, Wang, Kessel, Chen, \&
  Lepping}]{chao2002models}
Chao, J., Wu, D., Lin, C.-H., {et~al.} 2002, in Cospar Colloquia series,
  Vol.~12, Elsevier, 127--135

\bibitem[{Chasapis {et~al.}(2017)Chasapis, Matthaeus, Parashar, LeContel,
  Retin{\`o}, Breuillard, Khotyaintsev, Vaivads, Lavraud, Eriksson,
  {et~al.}}]{chasapis2017electron}
Chasapis, A., Matthaeus, W., Parashar, T., {et~al.} 2017, The Astrophysical
  Journal, 836, 247

\bibitem[{Chasapis {et~al.}(2018)Chasapis, Matthaeus, Parashar, Wan, Haggerty,
  Pollock, Giles, Paterson, Dorelli, Gershman, {et~al.}}]{chasapis2018situ}
---. 2018, The Astrophysical Journal Letters, 856, L19

\bibitem[{Chen {et~al.}(2019)Chen, Klein, \& Howes}]{chen2019evidence}
Chen, C., Klein, K., \& Howes, G.~G. 2019, Nature communications, 10, 1

\bibitem[{Drake {et~al.}(2003)Drake, Swisdak, Cattell, Shay, Rogers, \&
  Zeiler}]{drake2003formation}
Drake, J., Swisdak, M., Cattell, C., {et~al.} 2003, Science, 299, 873

\bibitem[{Ergun {et~al.}(2020)Ergun, Ahmadi, Kromyda, Schwartz, Chasapis,
  Hoilijoki, Wilder, Stawarz, Goodrich, Turner,
  {et~al.}}]{ergun2020observations}
Ergun, R., Ahmadi, N., Kromyda, L., {et~al.} 2020, The Astrophysical Journal,
  898, 154

\bibitem[{Fu {et~al.}(2016)Fu, Cao, Vaivads, Khotyaintsev, Andre, Dunlop, Liu,
  Lu, Huang, Ma, {et~al.}}]{Fu2016}
Fu, H., Cao, J., Vaivads, A., {et~al.} 2016, Journal of Geophysical Research:
  Space Physics, 121, 1263

\bibitem[{Gary {et~al.}(1993)Gary, Fuselier, \& Anderson}]{gary1993ion}
Gary, S.~P., Fuselier, S.~A., \& Anderson, B.~J. 1993, Journal of Geophysical
  Research: Space Physics, 98, 1481

\bibitem[{Gosling(2007)}]{gosling2007observations}
Gosling, J. 2007, The Astrophysical Journal Letters, 671, L73

\bibitem[{Greco {et~al.}(2008)Greco, Chuychai, Matthaeus, Servidio, \&
  Dmitruk}]{greco2008intermittent}
Greco, A., Chuychai, P., Matthaeus, W., Servidio, S., \& Dmitruk, P. 2008,
  Geophysical Research Letters, 35

\bibitem[{Greco {et~al.}(2009)Greco, Matthaeus, Servidio, Chuychai, \&
  Dmitruk}]{greco2009statistical}
Greco, A., Matthaeus, W., Servidio, S., Chuychai, P., \& Dmitruk, P. 2009, The
  Astrophysical Journal Letters, 691, L111

\bibitem[{He {et~al.}(2020)He, Zhu, Verscharen, Duan, Zhao, \&
  Wang}]{he2020spectra}
He, J., Zhu, X., Verscharen, D., {et~al.} 2020, The Astrophysical Journal, 898,
  43

\bibitem[{He {et~al.}(2019)He, Duan, Wang, Zhu, Li, Verscharen, Wang, Tu,
  Khotyaintsev, Le, {et~al.}}]{he2019direct}
He, J., Duan, D., Wang, T., {et~al.} 2019, The Astrophysical Journal, 880, 121

\bibitem[{Hollweg \& Isenberg(2002)}]{hollweg2002generation}
Hollweg, J.~V., \& Isenberg, P.~A. 2002, Journal of Geophysical Research: Space
  Physics, 107, SSH

\bibitem[{Huang {et~al.}(2018)Huang, Sahraoui, Yuan, Le~Contel, Breuillard, He,
  Zhao, Fu, Zhou, Deng, {et~al.}}]{huang2018observations}
Huang, S., Sahraoui, F., Yuan, Z., {et~al.} 2018, The Astrophysical Journal,
  861, 29

\bibitem[{Isenberg \& Vasquez(2019)}]{isenberg2019perpendicular}
Isenberg, P.~A., \& Vasquez, B.~J. 2019, The Astrophysical Journal, 887, 63

\bibitem[{Marsch \& Tu(1997)}]{marsch1997intermittency}
Marsch, E., \& Tu, C.-Y. 1997

\bibitem[{Osman {et~al.}(2014)Osman, Matthaeus, Gosling, Greco, Servidio, Hnat,
  Chapman, \& Phan}]{osman2014magnetic}
Osman, K., Matthaeus, W., Gosling, J., {et~al.} 2014, Physical Review Letters,
  112, 215002

\bibitem[{Osman {et~al.}(2012)Osman, Matthaeus, Hnat, \&
  Chapman}]{osman2012kinetic}
Osman, K., Matthaeus, W., Hnat, B., \& Chapman, S. 2012, Physical review
  letters, 108, 261103

\bibitem[{Phan {et~al.}(2007)Phan, Paschmann, Twitty, Mozer, Gosling, Eastwood,
  {\O}ieroset, Reme, \& Lucek}]{phan2007evidence}
Phan, T., Paschmann, G., Twitty, C., {et~al.} 2007, Geophysical Research
  Letters, 34

\bibitem[{Phan {et~al.}(2018)Phan, Eastwood, Shay, Drake, Sonnerup, Fujimoto,
  Cassak, {\O}ieroset, Burch, Torbert, {et~al.}}]{phan2018electron}
Phan, T., Eastwood, J.~P., Shay, M., {et~al.} 2018, Nature, 557, 202

\bibitem[{Pollock {et~al.}(2016)Pollock, Moore, Jacques, Burch, Gliese, Saito,
  Omoto, Avanov, Barrie, Coffey, {et~al.}}]{pollock2016fast}
Pollock, C., Moore, T., Jacques, A., {et~al.} 2016, Space Science Reviews, 199,
  331

\bibitem[{Priest(1986)}]{priest1986magnetic}
Priest, E. 1986, ppm, 1

\bibitem[{Retin{\`o} {et~al.}(2007)Retin{\`o}, Sundkvist, Vaivads, Mozer,
  Andr{\'e}, \& Owen}]{retino2007situ}
Retin{\`o}, A., Sundkvist, D., Vaivads, A., {et~al.} 2007, Nature Physics, 3,
  235

\bibitem[{Russell {et~al.}(2016)Russell, Anderson, Baumjohann, Bromund,
  Dearborn, Fischer, Le, Leinweber, Leneman, Magnes,
  {et~al.}}]{russell2016magnetospheric}
Russell, C., Anderson, B., Baumjohann, W., {et~al.} 2016, Space Science
  Reviews, 199, 189

\bibitem[{Schwartz {et~al.}(1996)Schwartz, Burgess, \& Moses}]{schwartz1996low}
Schwartz, S., Burgess, D., \& Moses, J. 1996, Annales Geophysicae, 14, 1134

\bibitem[{Servidio {et~al.}(2011)Servidio, Greco, Matthaeus, Osman, \&
  Dmitruk}]{servidio2011statistical}
Servidio, S., Greco, A., Matthaeus, W., Osman, K., \& Dmitruk, P. 2011, Journal
  of Geophysical Research: Space Physics, 116

\bibitem[{Shay {et~al.}(2014)Shay, Haggerty, Phan, Drake, Cassak, Wu, Oieroset,
  Swisdak, \& Malakit}]{shay2014electron}
Shay, M., Haggerty, C., Phan, T., {et~al.} 2014, Physics of Plasmas, 21, 122902

\bibitem[{Sundkvist {et~al.}(2007)Sundkvist, Retin{\`o}, Vaivads, \&
  Bale}]{sundkvist2007dissipation}
Sundkvist, D., Retin{\`o}, A., Vaivads, A., \& Bale, S.~D. 2007, Physical
  review letters, 99, 025004

\bibitem[{Torbert {et~al.}(2016)Torbert, Russell, Magnes, Ergun, Lindqvist,
  LeContel, Vaith, Macri, Myers, Rau, {et~al.}}]{torbert2016fields}
Torbert, R., Russell, C., Magnes, W., {et~al.} 2016, Space Science Reviews,
  199, 105

\bibitem[{Wan {et~al.}(2016)Wan, Matthaeus, Roytershteyn, Parashar, Wu, \&
  Karimabadi}]{wan2016intermittency}
Wan, M., Matthaeus, W., Roytershteyn, V., {et~al.} 2016, Physics of Plasmas,
  23, 042307

\bibitem[{Wang {et~al.}(2016)Wang, Lu, Nakamura, Huang, Du, Guo, Teh, Wu, Lu,
  \& Wang}]{wang2016coalescence}
Wang, R., Lu, Q., Nakamura, R., {et~al.} 2016, Nature Physics, 12, 263

\bibitem[{Wang {et~al.}(2019)Wang, Alexandrova, Perrone, Dunlop, Dong, Bingham,
  Khotyaintsev, Russell, Giles, Torbert, {et~al.}}]{wang2019vortex}
Wang, T., Alexandrova, O., Perrone, D., {et~al.} 2019, The Astrophysical
  Journal Letters, 871, L22

\bibitem[{Wang {et~al.}(2013)Wang, Tu, He, Marsch, \&
  Wang}]{wang2013intermittent}
Wang, X., Tu, C., He, J., Marsch, E., \& Wang, L. 2013, The Astrophysical
  Journal Letters, 772, L14

\bibitem[{Wilder {et~al.}(2018)Wilder, Ergun, Burch, Ahmadi, Eriksson, Phan,
  Goodrich, Shuster, Rager, Torbert, {et~al.}}]{wilder2018role}
Wilder, F., Ergun, R., Burch, J., {et~al.} 2018, Journal of Geophysical
  Research: Space Physics, 123, 6533

\bibitem[{Zhang {et~al.}(2015)Zhang, He, Tu, Yang, Wang, Marsch, \&
  Wang}]{zhang2015occurrence}
Zhang, L., He, J., Tu, C., {et~al.} 2015, The Astrophysical Journal Letters,
  804, L43

\bibitem[{Zhou {et~al.}(2017)Zhou, Berchem, Walker, El-Alaoui, Deng, Cazzola,
  Lapenta, Goldstein, Paterson, Pang, {et~al.}}]{zhou2017coalescence}
Zhou, M., Berchem, J., Walker, R., {et~al.} 2017, Physical review letters, 119,
  055101

\end{thebibliography}

\end{document}